\documentclass[english,twoside,a4paper,10pt]{article}

\usepackage[latin1]{inputenc}
\usepackage[T1]{fontenc}
\usepackage{amsmath}
\usepackage{amsfonts}
\usepackage{graphicx}
\usepackage{subfigure}
\usepackage{a4wide}
\usepackage{amssymb}
\usepackage{fancyhdr}
\usepackage{mathrsfs}
\usepackage[toc,page]{appendix}

\begin{document}

\title{\bf Reconstruction of inflation from scalar field non-minimally coupled with the Gauss-Bonnet term}
\author{ 
Lorenzo Sebastiani$^{1,2}$\footnote{E-mail: lorenzo.sebastiani@unitn.it
},\,\,\,
Shynaray Myrzakul$^{3}$\footnote{E-mail: shynaray1981@gmail.com},\,\,\,
Ratbay Myrzakulov$^{1}$\footnote{E-mail: rmyrzakulov@gmail.com}\\
\\
\begin{small}
$^{1}$ Department of General \& Theoretical Physics and Eurasian Center for
\end{small}\\
\begin{small} 
Theoretical Physics, Eurasian National University, Astana 010008, Kazakhstan
\end{small}\\
\begin{small}
$^2$ Dipartimento di Fisica, Universit\`a di Trento, Italy 
\end{small}\\
\begin{small}
$^3$ Department of Theoretical and Nuclear Physics,
\end{small}\\
\begin{small}
Al-Farabi Kazakh National University, Al-Farabi Almaty, Kazakhstan
\end{small}
}

\date{}

\maketitle

%%%%%%%%%%%%%%%%%%%%%%%%%%%%%%%%%%%%%%%%%%%%%%%%%%%%%%%%%%%%%%%%%%%%%%%%%%%%%%%%%%%%%%%%%%%%%%%%%%%%%%%%%%%%%%%%%%%%%%%%%%%%%%%%%%%

%%%%%%%%%%%%%%%%%%%%%
%  Abstract
%%%%%%%%%%%%%%%%%%%%%
\begin{abstract}
In this paper, we analyze the early-time inflation in a scalar-tensor theory of gravity where the scalar field is minimally coupled with the 
Gauss-Bonnet four dimensional topological invariant. The theory belongs to a class of Horndeski models where the field equations are at the second order like in 
General Relativity. A viable inflationary scenario must correctly reproduce the last Plank satellite data. By starting from some simple assumptions on the field and on the coupling 
function between the field and the Gauss-Bonnet term, we derive the spectral index and the tensor-to-scalar ratio of the model. Once the model is viable, it is finally possible
to fully reconstruct its Lagrangian.
\end{abstract}
%%%%%%%%%%%%%%%%%%%%%

%----------------------------
%PACS
%----------------------------

%===========================================================================

\tableofcontents
%%%%%%%%%%%%%%%%%%%%%%%%%%%
%%%  Sec. I
%%%%%%%%%%%%%%%%%%%%%%%%%%%
\section{Introduction}

In the last three decedes, after the proposal of the inflationary paradigm by Guth~\cite{Guth} and Sato~\cite{Sato}, several scenarios for describing the early-time accelerated expansion of our Universe have been carried out (see Refs.~\cite{Linde, revinflazione} for some reviews). Among them, the ``old inflationary scenario'' is based on canonical scalar field theories where a scalar field, dubbed ``inflaton'', drives the primordial acceleration. Later, new classes  of scalar theories have been proposed, like the $k$-essence models~\cite{kess1, kess2, kess3}, where the Lagrangian of the field contains higher order kinetic terms and which lead to the suppression of the speed of the sound: as a consequence, the tensor-to-scalar ratio associated to the tensorial cosmological perturbations is extremelly small, as it is strongly encouraged by the cosmological data~\cite{Planck}.

An alternative description of inflation can be furnished by scalar-tensor theories of gravity, where a scalar field is coupled with some curvature invariants (Einstein's tensor, Ricci scalar...) inside the gravitational action. In this kind of theories the field equations are higher order differential equations, but in 
1974 Horndeski derived the most general class of scalar-tensor models which lead to second order differential equations like in the theory of Einstein~\cite{Horn}. 
Recently, inflation from Horndeski gravity has been investigated in several works
~\cite{Amendola, Def, DeTsu, DeFelice, Kob, Kob2, Qiu, EugeniaH, mioH, mioGB}.
In Ref.~\cite{Staro1} an investigation on
the homogeneous and isotropic cosmologies in some classes of models of Horndeski gravity
with Galileon shift symmetry has been carried out.

In this paper, we will consider a class of Horndeski Lagrangian where a scalar $k$-essence field supporting 
inflation is coupled with the Gauss-Bonnet term. We mention that modifications of gravity from the four dimensional Gauss-Bonnet
topological invariant have been often considered in the context of high curvature corrections of General Relativity as the result of quantum
gravity effects  (see for example Refs.~\cite{RGinfl, r2} or Refs.~\cite{GBO1}--\cite{GB03}). We will follow the lines of Refs.~\cite{muk1, muk2, miorec, miorec2} and we will propose a reconstruction method in order to infer viable models in agreement with the cosmological data. In this respect, we note that one of the most robust predictions of inflation is represented by the possibility of reproducing the cosmological perturbations at the origin of the inhomogeneities of our Friedmann Universe. The last Planck satellite data lead to a spectral index $n_s\simeq 1-2/N$ and to a tensor-to-scalar ratio $r< 8/N$ or $r\sim 1/N^2$, where the $e$-folds number $N$ must be $N\simeq 60$ in order to explain the thermalization of our observable Universe. By starting from some simple Ansatz 
on the $k$-essence field and on the coupling function of the field with the Gauss-Bonnet, it
is possible to derive these indexes and finally get the full form of the viable models. 

The paper is organized as follows. In Section {\bf 1} we present the model of $k$-essence coupled with the Gauss-Bonnet in the Horndeski framework. In Section {\bf 2} we study the background equations for inflation and the equations for cosmological perturbations, deriving the spectral index and the tensor-to-scalar ratio. In Section {\bf 4} we introduce our Ansatz for the scalar field and the coupling function between the field and the Gauss-Bonnet term. Thus, in Section {\bf 5} we reconstruct several inflationary viable models of canonical scalar field, while in Section {\bf 6} the case of $k$-essence with quadratic kinetic term is investigated. Conclusions and final remarks are given in Section {\bf 7}.

%%% Unit %%%
We use units of $k_{\mathrm{B}} = c = \hbar = 1$ and 
$8\pi/M_{Pl}^2=1$, where $M_{Pl}$ is the Planck Mass.
%%%%%%%%%%%%
%%%%%

\section{The model}

In this paper we will work with the following gravitational model,
\begin{equation}
I=\int_\mathcal M dx^4\sqrt{-g}\left[\frac{R}{2}+\xi(\phi) \mathcal G+p(\phi, X)\right]\,,
\label{action}
\end{equation}
where $\mathcal M$ is the space-time manifold and $g$ is the determinant of the metric tensor $g_{\mu\nu}$. The Hilbert-Einstein action of General Relativity (GR), given by the Ricci scalar  $R$, has been modified by introducing a coupling $\xi(\phi)$ between a scalar field $\phi$ and the Gauss-Bonnet four dimensional topological invariant  $\mathcal G$,
\begin{equation}
\mathcal G=R^2-4R_{\mu\nu}R^{\mu\nu}+R_{\mu\nu\sigma\xi}R^{\mu\nu\sigma\xi}\,,
\end{equation}
with $R_{\mu\nu}$ and $R_{\mu\nu\sigma\xi}$ the Ricci tensor and the Riemann tensor, respectively. Finally, 
 $p(\phi, X)$ is a function of the scalar field $\phi$ and its kinetic energy $X$, 
\begin{equation}
X=-\frac{g^{\mu\nu}\partial_\mu \phi\partial_\nu\phi}{2}\,.
\end{equation}
The scalar field effective pressure corresponds to the field Lagrangian $p(\phi, X)$, while the effective energy density of the field $\rho(\phi, X)$ is derived as
\begin{equation}
\rho(\phi, X)=2X\frac{\partial p(\phi, X)}{\partial X}-p(\phi, X)\,,\label{4}
\end{equation}
such that the following relation holds true:
\begin{equation}
\rho(\phi, X)+p(\phi, X)=2X p_X(\phi, X)\,.
\end{equation}
The case of canonical scalar field is given by $p(\phi, X)=X-V(\phi)$, $V(\phi)$ being a function of the field only, while in $k$-essence Lagrangian higher order kinetic terms appear~\cite{kess1, kess2}.

This kind of scalar-tensor model with a minimal coupling with the Gauss-Bonnet belongs to a subclass of Horndeski theories of gravity~\cite{Horn, Kob} and the field equations are at the second order like in the theory of Einstein.

%%%%%%%%%%%%%%%%%%%%%%%%%%%%%%%%%%%%%%%%%%%%%%%

\section{Inflation}

In this section we will study the inflationary cosmology with our model. At first, we will analyze the equations of the bulk, and then we will proceed with the study of cosmological perturbations.

\subsection{Background equations}

The flat Friedmann-Robertson-Walker (FRW) metric reads,
\begin{equation}
ds^2=-dt^2+a(t)^2 d{\bf x}^2\,,\label{metric}
\end{equation}
where $a(t)$ is the scale factor and depends on the cosmological time.
An useful parametrization of the field equations follows from the introduction of the $e$-folds number $N$ as,
\begin{equation}
N=\log\left[\frac{a(t_0)}{a(t)}\right]\,,\label{N}
\end{equation}
where $a(t_0)$ is the scale factor at the time $t_0$ when inflation ends, such that $t<t_0$ and $0<N$.

In terms of the $e$-folds the first Friedmann equation of the model (\ref{action}) leads to~\cite{mioGB},
\begin{equation}
3H^2=\rho(\phi, X)+
24H^4\phi'\frac{d\xi(\phi)}{d\phi}
\,,\label{EOM}
\end{equation}
while the conservation law reads
\begin{equation}
-\rho'(\phi, X)+3H^2\phi'^2(p_X(\phi, X))
=24\frac{d\xi(\phi)}{d\phi}\phi' H^3(H'-H)
\,,
\label{conslaw}
\end{equation}
where the prime denotes the derivative with respect to $N$ and $X=H^2\phi'^2/2$.

The early-time inflation takes place at high curvature and is described by a (quasi) de Sitter solution where the Hubble parameter is almost a constant: this is the so called slow-roll approximation regime. Therefore, the field slowly moves and drives the exit from the accelerated epoch.

During the the slow-roll regime the field evolves under the conditions
\begin{equation}
\phi'^2\ll 1\,,\quad |\phi''|\ll |\phi'|\,,
\end{equation}
and the $\epsilon$ slow-roll parameter
\begin{equation}
\epsilon=\frac{H'}{H}\,,\label{epsilon}
\end{equation}
is positive and very small. Acceleration ends when $\epsilon$ is on the order of the unit.
By taking into account the slow-roll approximation, equations (\ref{EOM})--(\ref{conslaw}) read
\begin{equation}
3H^2\simeq \rho(\phi, X)\,,\label{uno}
\end{equation}
\begin{equation}
\rho'(\phi, X)-3H^2 \phi'^2 p_X(\phi, X)\simeq 
24\frac{d\xi(\phi)}{d\phi}\phi' H^4
\,.\label{due}
\end{equation}
These equations describe the Hubble parameter and the evolution of the field during inflation.

%%%%%%%%%%%%%%%%%%%%%%%%%%%%%%%%%%%%

\subsection{Cosmological perturbations}

Scalar metric perturbations around the FRW metric (\ref{metric}) in their general formulation read~\cite{Def, DeTsu, DeFelice},
\begin{equation}
ds^2=-[(1+\alpha(t, {\bf x}))^2-a(t)^{-2}\text{e}^{-2\zeta(t, {\bf x})}(\partial \psi(t,{\bf x}))^2]dt^2+2\partial_i\psi
(t,{\bf x})dt dx^i+a(t)^2
\text{e}^{2\zeta(t, {\bf x})}d{\bf x}\,,
\end{equation}
with $\alpha\equiv\alpha(t, {\bf x})\,,\psi\equiv\psi(t, {\bf x})$ and $\zeta\equiv\zeta(t,{\bf x})$ functions of the space-time coordinates. 

Thus, by
using the relations between $\alpha\,,\psi\,,\zeta$ that follow from the field equations, 
 the second-order action for perturbations reduces to~\cite{DeTsu, DeFelice},
\begin{equation}
I=\int_\mathcal{M}dx^4 a^3 Q\left[\dot\zeta^2-\frac{c_s^2}{a^2}(\nabla\zeta)^2\right]\,,\label{pertaction}
\end{equation}
where, if one uses the slow-roll approximation and in terms of the $e$-folds number, 
\begin{eqnarray}
Q =
\frac{\phi'^2}{2H^2}
\left(
96H^4\xi'^2(\phi)+p_X(\phi, X)+\phi'^2 p_{XX}(\phi, X)
\right)\,,\label{Q}
\end{eqnarray}
while the
square of the speed of sound reads
\begin{equation}
\hspace{-2cm}
c_s^2=
\frac{p_X(\phi, X)}{p_X(\phi, X)+96H^4\xi_\phi^2(\phi)+2p_{XX}(\phi, X)X}
\,.\label{c2}
\end{equation}
This last quantity plays a fundamental role in the evolution of cosmological perturbations. 
We note that, even in the case of canonical scalar field with $p_{XX}(\phi, X)=0$, one finds $c_s^2<1$, thanks to the contribute of the Gauss-Bonnet. This result seems to improve the predictivity of the model with respect to the standard framework of GR or other Horndeski theories (see for example Ref.~\cite{miorec2}), due to the fact that, when $c_s^2$ is small, the tensor-to-scalar ratio of tensorial perturbations tends to be suppressed, in agreement with cosmological data.

The action for scalar perturbations can be rewritten as,
\begin{equation}
I=\int dx^4\left[\dot v^2-\frac{c_s^2}{a^2}(\nabla v)^2+\ddot z\frac{v^2}{z}\right]\,,\label{action2}
\end{equation}
where,
\begin{equation}
v\equiv v(t, {\bf x})=z(t) \zeta(t, {\bf x})\,,\quad z\equiv z(t)=\sqrt{a^3 Q}\,.
\end{equation}
As a consequence, the field equation for perturbations is derived as,
\begin{equation}
\ddot v-\frac{c_s^2}{a^2}\bigtriangleup v-\frac{\ddot z}{z}v=0\,.
\end{equation}
Therefore, if one decomposes  
$v(t, {\bf x})$ in Fourier modes $v_k\equiv v_k(t)\exp[i {\bf k}{\bf x}]$, we obtain
\begin{equation}
\ddot v_k+\left(k^2\frac{c_s^2}{a^2}-\frac{\ddot z}{z}\right)v_k=0\,.\label{eqpert}
\end{equation}
The solution of this equation, back into the asymptotic past, leads to
\begin{equation}
\zeta_k\equiv \frac{v_k}{\sqrt{Q a^3}}\simeq i\frac{H}{2\sqrt{Q}(c_s k)^{3/2}}
\text{e}^{\pm i k\int \frac{c_s}{a}dt}
\left(1+i c_s k\int\frac{dt}{a}\right)\,.
\end{equation}
Now we can calculate 
the variance of the power spectrum of perturbations on the sound horizon crossing $c_s\kappa\simeq H a$, namely
\begin{equation}
\mathcal P_{\mathcal R}\equiv\frac{|\zeta_k|^2 k^3}{2\pi^2}|_{c_s k\simeq H a}=\frac{H^2}{8\pi^2 c_s^3 Q}|_{c_s k\simeq H a}\,.
\end{equation}
From the variance of the power spectrum one gets the spectral index $n_s$~\cite{mioGB},
\begin{eqnarray}
(1-n_s)&=&-\frac{d\ln \mathcal P_{\mathcal R}}{d \ln k}|_{k=a H/c_s}\nonumber\\
&=&
\left(\phi ' \left(3072 H^6 p_X \xi '^3+24 H^4 \xi
   ' \left(p_X \left(16 p_X \xi ' \phi '^2+8
   \xi ''+p_{XX}\phi '^4\right)-12 \xi '
   p_X'\right)
\right.\right.
\nonumber\\&&
+H^2 \left(16 p_X^2 \xi ' \phi
   '^2+3 p_X^2 p_{XX} \phi
   '^6+\phi '^4 \left(p_X p_{XX}'-3
   p_{XX}p_X'\right)\right)
\nonumber\\&&
\left.\left.
+2 p_X^3 \phi
   '^4-2 p_X p_X' \phi '^2\right)-2
   p_X \phi '' \left(288 H^4 \xi '^2+H^2 p_{XX}
   \phi '^4+2 p_X \phi '^2\right)\right)
\nonumber\\&&
\times\frac{1}{2
   p_X \phi ' \left(96 H^4 \xi '^2+H^2 p_{XX}
   \phi '^4+p_X \phi '^2\right)}\,,\label{n}
\end{eqnarray}
where we used (\ref{Q})--(\ref{c2}).

In a similar way it is possible to calculate the tensor-to-scalar ratio for the tensorial perturbations, 
\begin{equation}
r\simeq
\frac{8 p_X\phi '^2
   \sqrt{\frac{p_X}{\frac{96 H^4 \xi
   '^2}{\phi '^2}+H^2 p_{XX} \phi
   '^2+p_X}}}{\frac{4 H^2 \left(2-\log \left(
   H^2 \phi '^2/2\right)\right) \left(\xi '' \phi '-\xi '
   \phi ''\right)}{\phi '}+1}\,.\label{r}
\end{equation}
These indexes describe the cosmological perturbations left at the end of inflation and must be evaluated at the beginning of the eraly-time acceleration when $N$ has to be $N\simeq 60$.

\section{Viable models for inflation}

Inflation predicts the production of cosmological perturbations responsable of anisotropies of our Universe at the galactic scale. The last Planck satellite data~\cite{Planck} fit the spectral index and the tensor-to-scalar ratio as
$n_{\mathrm{s}} = 0.968 \pm 0.006\, (68\%\,\mathrm{CL})$ and 
$r < 0.11\, (95\%\,\mathrm{CL})$. Thus, by taking into account that $N\simeq 60$, 
observations strongly encourage the models with
$(1-n_s)\simeq 2/N$ and 
$r< 8/N$
or
$r\sim1/N^2$.

In order to reconstruct some viable models we will start from the following simple assumptions on the background solutions of the scalar field and of the coupling function between the field and the Gauss-Bonnet term,
\begin{equation}
\phi'^2=\frac{\alpha^2}{N^a}\,,\quad \alpha<0\,,1\leq a\,,\label{Anphi}
\end{equation}
\begin{equation}
\xi(\phi)=\frac{\beta}{N^b}\,,\quad 1\leq b\,,\label{Anxi}
\end{equation}
where $\alpha\,,\beta$ are dimensional constants and $a, b$ are positive parameters larger or equal to one. 

For the Lagrangian of the field $p(\phi, X)$ we will assume
\begin{equation}
p(\phi, X)=\kappa X^\lambda-V(\phi)\,,\quad 0<\lambda\,,\label{pk}
\end{equation}
with $\lambda$ a positive parameter, $\kappa$ a (positive) dimensional constant and $V(\phi)$ a function of the field.
Therefore, the effective field energy density (\ref{4}) leads to
\begin{equation}
\rho(\phi, X)=\kappa(2\lambda-1)X^\lambda+V(\phi)\,,\label{rhok}
\end{equation}
and Eqs.(\ref{uno})--(\ref{due}) in the slow-roll approximation can be rewritten as
\begin{equation}
H^2\simeq\frac{V(N)}{3}\,,\label{unobis}
\end{equation}
\begin{equation}
-6\kappa\lambda X^{\lambda}
\simeq 24\xi' H^4-V'(N)\,.\label{duebis}
\end{equation}
By starting from this equations with the Ansatz in (\ref{Anphi})--(\ref{Anxi}), we can find the on-shell potential and the Hubble parameter for different kinds of model. As a consequence, we can derive the spectral index (\ref{n}) and the tensor-to-scalar ratio (\ref{r}) in order to study the viable conditions of the theory. Finally, an explicit reconstruction for the coupling function $\xi(\phi)$ and for the Lagrangian of the field is possible by inverting the relation (\ref{Anphi}).

In the following sections we will analyze some examples.

\section{Canonical scalar field}

In this section, we will consider canonical scalar field models where  $\kappa=\lambda=1$ in (\ref{pk})--(\ref{rhok}).

Let us start with the case $a=1$ in (\ref{Anphi}). It follows,
\begin{equation}
\phi=2\alpha\sqrt{N}\,,\label{phi1}
\end{equation}
where we remember that $\alpha<0$.
From (\ref{duebis}) we can obtain the following on shell form of the potential,
\begin{equation}
V(N)=\frac{3N^b(\alpha^2-b)}{8b\beta}\,.
\end{equation}
Therefore, by plugging this expression in (\ref{unobis}) one has for the Hubble parameter:
\begin{equation}
H^2\simeq\frac{N^b(\alpha^2-b)}{8b\beta}\,.
\end{equation}
The $\epsilon$ slow-roll parameter (\ref{epsilon}) describing the evolution of the quasi-de Sitter universe is given by,
\begin{equation}
\epsilon\simeq\frac{b}{2N}\,.
\end{equation}
The square of the speed of the sound (\ref{c2}) reads,
\begin{equation}
c_s^2\simeq\frac{1}{1+\frac{3(b-\alpha^2)^2}{2\alpha^2 N}}\,,
\end{equation}
and we see that, thanks to the non-minimal coupling with the Gauss-Bonnet, it is smaller than one, even if it is extremelly close to one when $1\ll N$.

Now we can derive the spectral index (\ref{n}) and the tensor-to-scalar ratio (\ref{r}),
\begin{equation}
(1-n_s)\simeq \frac{1+(b-\alpha^2)}{N}\,,\quad r\simeq \frac{8\alpha^2}{N}\,.
\end{equation}
Thus, in order to satisfy the last Planck satellite data we must require
\begin{equation}
(b-\alpha^2)=1\,,\quad\alpha^2<1\,.\label{cond1}
\end{equation}
As a consequence, $\beta$ must be negative to get a real solution for the Hubble parameter. By using (\ref{phi1}) one has
\begin{equation}
N=\frac{\phi^2}{4\alpha^2}\,,
\end{equation}
and the model is fully reconstructed as 
\begin{equation}
\xi(\phi)=\frac{(4\alpha^2)^b\beta}{\phi^{2b}}
\,,\quad
V(\phi)=-\frac{3}{2^{3+2b}b\beta}\left(\frac{\phi^2}{\alpha^2}\right)^b\,,\quad 1\leq b<2\,, \beta<0\,,
\end{equation}
where we used (\ref{cond1}).
We should point out that canonical scalar fields with power law  potential in the framework of General Relativty do not leave to viable scenarios for inflation. In particular the quadratic potential, namely one of the first models of the ``old inflationary scenario'', leads to a viable spectral index, but the tensor-to-scalar ratio reads $r\simeq 8/N$ and results to be large. Here, in the framework of Horndeski gravity, power-law potentials $V(\phi)\sim \phi^q$, $2\leq q<4$, bring to models compatible with observations.\\
\\
Let us take now the case $a=2$ in (\ref{Anphi}), namely  
\begin{equation}
\phi=\phi_\text{i}+\alpha\log[N/N_\text{i}]\,,\label{phi2}
\end{equation}
where $\phi_\text{i}<0$ is the value of the field at the beginning of inflation when $N=N_\text{i}$. 
The field potential is derived from (\ref{duebis}) as
\begin{equation}
V(N)=\frac{3\alpha^{2b}\text{e}^{-\frac{\alpha^2}{N}}}{V_0+8b\beta\Gamma[b,\frac{\alpha^2}{N}]}\,,\label{42}
\end{equation}
where $V_0$ is a constant and $\Gamma[b, \frac{\alpha^2}{N}]$ corresponds to the upper incomplete gamma function,
\begin{equation}
\Gamma\left[b, \frac{\alpha^2}{N}\right]=\int^\infty_{\alpha^2/N} t^{b-1}\text{e}^{-t}dt\,.
\label{Gamma}
\end{equation}
As a consequence, the Hubble parameter during inflation reads:
\begin{equation}
H^2\simeq \frac{\alpha^{2b}}{V_0+8b\beta\Gamma[b,0]}\,.
\end{equation}
In the limit $0\ll N$ the $\epsilon$ slow-roll
parameter is given by (we remember $1\leq b$),
\begin{eqnarray}
\epsilon&\simeq&\frac{\alpha^2 V_0}{2(V_0+8\beta)N^2}\,,\quad b=1\,,\nonumber\\
\epsilon&\simeq& \frac{\alpha^2}{2N^2}\,,\quad 1<b\,.
\end{eqnarray}
The square of the speed of sound results to be $c_s\simeq 1^-$. Moreover, for the spectral index and the tensor-to-scalar ratio one has:
\begin{equation}
(1-n_s)\simeq\frac{2}{N}\,,\quad r\simeq\frac{8\alpha^2}{N^2}\,.
\end{equation}
As a general result, we can say that this kind of model is viable and correctly reproduce the last Planck satellite data when $\alpha^2<0$. The potential and the coupling function are reconstructed as,
\begin{equation}
V(\phi)=\frac{3\alpha^{2b}\text{e}^{-\frac{\alpha^2\exp[(\phi_0-\phi)/\alpha]}{N_\text{i}}}}{V_0+8b\beta\Gamma[b,\frac{\alpha^2\exp[(\phi_0-\phi)/\alpha]}{N_\text{i}}]}\,,
\quad 
\xi(\phi)=\beta\frac{\text{e}^{-b(\phi-\phi_0)/\alpha}}{\left(N_\text{i}\right)^b}\,.
\end{equation}
For example, when $b=1$ we get $V(\phi)=3\alpha^2/\left[8\beta+V_0\exp\left[(\alpha^2/N_\text{i})\text{e}^{(\phi_0-\phi)/
\alpha}\right]\right]$.\\
\\
It is also possible to investigate the model for generic parameters of $a$ and $b$ in (\ref{Anphi})--(\ref{Anxi}). From (\ref{duebis}) we can get
\begin{equation}
V(N)=\frac{3(a-1)\text{e}^{\frac{\alpha^2 N^{1-a}}{1-a}}}{8b\beta\Gamma[\frac{b}{a-1},\frac{\alpha^2 N^{1-a}}{a-1}]}\left(\frac{\alpha^2}{a-1}\right)^{b/(a-1)}\,,\quad a\neq 1\,,
\end{equation}
where we must require
\begin{equation}
a-1\leq b\,,
\end{equation}
in order to obtain a finite value for the incomplete gamma function (\ref{Gamma}) when $0\ll N$. By making use of the relation $H^2=V(\phi)/3$, it is possible to calculate the spectral index $n_s$ (\ref{n}), namely
\begin{equation}
n_s\simeq \frac{a}{N}\,,
\end{equation}
and we see that only the models with $a=2$ are viable, namely we recover the example in (\ref{42}).

\section{$k$-essence with quadratic kinetic term}

In this section, we will generalize our analysis to $k$-essence models with quadratic kinetic term, namely $\lambda=2$ in (\ref{pk})--(\ref{rhok}). We start with $a=1$ in (\ref{Anphi}), namely we assume (\ref{phi1}). Thus, from (\ref{duebis}) one may get,
\begin{equation}
V(N)=\frac{3N^{1+b}}{\alpha^4\kappa N^b-8\beta N+V_0 N^{1+b}}\,,\label{51}
\end{equation}
with $V_0$ constant, such that
\begin{equation}
H^2\simeq\frac{N^{1+b}}{\alpha^4\kappa N^b-8\beta N+V_0 N^{1+b}}\,.
\end{equation}
Note that if $V_0\neq 0$ the Hubble parameter behaves as $H^2\simeq 1/V_0$, but for $V_0=0$ other forms of the Hubble parameter are allowed. The $\epsilon$ slow-roll parameter reads,
\begin{equation}
\epsilon\simeq\frac{\alpha^4\kappa N^b-8b\beta N}{2N(V_0 N^{1+b}-8\beta N+\alpha^4\kappa N^b)}\,.
\end{equation}
Thus,
\begin{eqnarray}
\epsilon&\simeq&\frac{1}{2N}\,,\quad V_0=0\,,
\nonumber\\
\epsilon&\simeq& \frac{\alpha^4\kappa-8b\beta N^{1-b}}{2V_0 N^2}\,,\quad V_0\neq 0\,,
\end{eqnarray}
bringing to different dynamics for the early-time expansion. In both of the cases,
\begin{equation}
c_s\simeq\frac{1}{3}\,,
\end{equation}
and the results for the spectral index (\ref{n}) are,
\begin{eqnarray}
(1-n_s)&\simeq&\frac{48\beta-5\kappa\alpha^4}{(24\beta-3\alpha^4\kappa)N}\,,\quad
V_0=0\,, b=1\,,
\nonumber\\
(1-n_s)&\simeq&\frac{5}{3N}\,,\quad V_0=0\,,1<b\,,
\nonumber\\
(1-n_s)&\simeq& \frac{2}{N}\,,\quad V_0\neq 0\,, 1\leq b\,.
\end{eqnarray}
The last two cases satisfy the last Planck satellite data, while in the first one we must require
\begin{equation}
\frac{48\beta-5\kappa\alpha^4}{(24\beta-3\alpha^4\kappa)}=2\,,\quad V_0=0\,, b=1\,.
\end{equation}
The tensor-to-scalar ratio (\ref{r}) reads
\begin{eqnarray}
r&\simeq&\frac{8\kappa\alpha^4}{\sqrt{3}(\alpha^4\kappa-8\beta)N}\,,\quad V_0=0\,,b=1\,,\nonumber\\
r&\simeq&\frac{8}{\sqrt{3}N}\,,\quad V_0=0\,, 1<b\,,\nonumber\\
r&\simeq&\frac{8\alpha^4\kappa}{\sqrt{3}V_0 N^2}\,,\quad V_0\neq 0\,, 1\leq b\,.
\end{eqnarray}
Thus, the model correctly reproduces also the tensorial cosmological perturbations (in the first case, we must require $\kappa\alpha^4/(\alpha^4\kappa-8\beta)\leq 1$).

The fully reconstruction of the potential leads to
\begin{equation}
V(\phi)=\frac{3\phi^2(\phi^2/\alpha^2)^b}{(\phi^2/\alpha^2)^b(4\alpha^6\kappa+V_0\phi^2)-2^{3+2b}\beta\phi^2}\,,\quad
\xi(\phi)=4^b\beta\left(\frac{\alpha^2}{\phi^2}\right)^b\,.
\end{equation}
For example, for $b=1$ and $V_0=0$, we recover a quadratic potential.\\
\\
Let us check for a more general result with generic $a\,,b$ in (\ref{Anphi})--(\ref{Anxi}). The potential reads,
\begin{equation}
V(N)=\frac{3(2a-1)N^{2a+b}}{\alpha^4\kappa N^{1+b}-8\beta(2a-1)N^{2a}-V_0 N^{2a+b}}\,,
\end{equation}
with $V_0$ constant. The Hubble parameter during the inflation is derived as (here, we remember $1\leq a,b$),
\begin{eqnarray}
H^2&\simeq&\frac{3(2a-1)N^{2a+b}}{\alpha^4\kappa N^{1+b}-8\beta(2a-1)N^{2a}}\,,\quad V_0= 0\,,\nonumber\\
H^2&\simeq&\frac{3-6a}{3V_0}\,,\quad V_0\neq 0\,.
\end{eqnarray}
The $\epsilon$ slow-roll parameter is given by
\begin{equation}
\epsilon\simeq\frac{(2a-1)(8b\beta N^{2a}-\alpha^4\kappa N^{1+b})N^b}{ 2(V_0 N^{2a+b}+8\beta(2a-1)N^{2a}-\alpha^4\kappa N^{1+b})}\,,
\end{equation}
and for $1\ll N$ we can verify that $\epsilon\ll 1$. The spectral index $n_s$ reads
\begin{equation}
(1-n_s)\simeq \frac{2a}{N}\,,\quad V_0\neq 0\,,1<a\leq b\,,
\end{equation}
We conclude that only the case $a=1$ is viable and we recover the model in (\ref{51}).

\section{Conclusions}

In this paper we analyzed an Horndeski model for inflation where the scalar field is non-minimally coupled with the Gauss-Bonnet four dimensional topological invariant.
Horndeski models are quite interesting since, despite the complexity of the Lagrangian, lead to second order differential equations like in General Relativity. Moreover, the four dimensional Gauss-Bonnet topological invariant is often analyzed in the context of higher curvature corrections of the Einstein's theory as a leading term from quantum corrections or string inspired theories. Finally, the scalar Horndeski field has been identified with a generic $k$-essence field, namely high order kinetic terms can appear in its Lagrangian.

Since a viable model for inflation must correctly reproduce the spectral index and the tensor-to-scalar ratio observed in our Universe, we reconstructed our models by starting from these indexes. To get them, we posed some Ansatz on the scalar field and on the coupling function between the field and the Gauss-Bonnet. In our analysis, we considered canonical scalar field and $k$-essence with quadratic kinetic term.

As a general observation, we can say that only when the derivative of the field with respect to the $e$-folds number behaves as $\phi'^2\sim 1/N$ or $\phi'^2\sim 1/N^2$ we get a viable scenario, confirming the result of Refs.~\cite{muk1, miorec} for the framework of General Relativity and of Ref.~\cite{miorec2} for Horndeski gravity with the coupling with the Einstein's tensor.

For inflation from quantum corrections to General Relativity see Ref.~\cite{buch}. Other works 
about modified gravity and inflation can be found in Refs.~\cite{Vagno, I2} or in Refs.~\cite{Odinfrev, FRreview}.

%%%%%%%%%%%%%%%%%%%%BIBLIO

\end{document}